\documentclass[twocolumn,pra,superscriptaddress,floatfix,showpacs]{revtex4}
\usepackage{amssymb}
\usepackage{amsmath}
\usepackage{graphicx}
\usepackage{amsfonts}

\begin{document}

\title{Coherent control of photon propagation \\
via electromagnetically induced transparency in lossless media}
\author{Liang He}
\affiliation{Institute of Theoretical Physics, Chinese Academy of Sciences, Beijing
100080, China}
\affiliation{Frontier Research System, The Institute of Physical and Chemical Research
(RIKEN), Wako-shi, Saitama 351-0198, Japan }
\author{Yu-xi Liu}
\affiliation{Frontier Research System, The Institute of Physical and Chemical Research
(RIKEN), Wako-shi, Saitama 351-0198, Japan }
\affiliation{CREST, Japan Science and Technology Agency (JST), Kawaguchi, Saitama
332-0012, Japan}
\author{S. Yi}
\affiliation{Institute of Theoretical Physics, Chinese Academy of Sciences, Beijing
100080, China}
\author{C. P. Sun}
\affiliation{Institute of Theoretical Physics, Chinese Academy of Sciences, Beijing
100080, China}
\affiliation{Frontier Research System, The Institute of Physical and Chemical Research
(RIKEN), Wako-shi, Saitama 351-0198, Japan }
\author{Franco Nori}
\affiliation{Frontier Research System, The Institute of Physical and Chemical Research
(RIKEN), Wako-shi, Saitama 351-0198, Japan }
\affiliation{CREST, Japan Science and Technology Agency (JST), Kawaguchi, Saitama
332-0012, Japan}
\affiliation{Center for Theoretical Physics, Physics Department, Center for the Study of
Complex Systems, The University of Michigan, Ann Arbor, Michigan 48109-1040,
USA}
\date{\today }

\begin{abstract}
We study the influence of a lossless material medium on the
coherent storage and quantum state transfer of a quantized probe
light in an ensemble of $\Lambda $-type atoms. The medium is
modeled as uniformly distributed two-level atoms with same energy
level spacing, coupling to a probe light. This coupled system can
be simplified to a collection of two-mode polaritons which couple
to one transition of the $\Lambda$-type atoms. We show that, when
the other transition of $\Lambda$-type atoms is controlled by a
classical light, the electromagnetically induced transparency can
also occur for the polaritons. In this case the coherent storage
and quantum transfer for photon states are achievable through the
novel dark states with respect to the polaritons. By calculating
the corresponding dispersion relation, we find the ensemble of the
three-level atoms with $\Lambda$-type transitions may serve as
quantum memory for it slows or even stops the light propagation
through the mechanism of electromagnetically induced transparency.
\end{abstract}

\maketitle

\section{Introduction}

Electromagnetically induced transparency (EIT) \cite{harris} is a typical
quantum coherent effect, in which the propagation of probe a field in a $%
\Lambda$-type atom ensemble can be well controlled by a classical light \cite%
{hau,Kash}. Most recently, the EIT phenomenon was suggested as an active
mechanism \cite{Lukin1,Lukin2,Lukin3} to slow down and even stop the photon
propagation, so that the photon state can be stored or released coherently.
These investigations \cite{Lukin1,Lukin2,Lukin3} are mainly motivated by the
fast development of quantum information science and technology \cite{Zei}.
This is because, with the help of quantum storage, one could complete a
series of quantum logical operations within the decoherence time.

In this paper we study the EIT mechanism for quantum information
processing in the presence of a lossless medium. This is motivated
by two reasons. Firstly, we notice that buffer gases, with
different atom species, are used in some of the recent EIT
experiments~\cite{buffer1, buffer2}. Usually one introduces a
buffer gas to lengthen the ground-state coherence lifetime of
confined EIT atoms. For the EIT effect in a $\Lambda $--sample
with a buffer gas, the probe field has a low group velocity when
it has a small detuning with respect to resonance \cite{buffer1,
buffer2}. These coherent phenomena essentially result from the
gaseous medium : the buffer gas. To see the coherent effect of the
buffer gas, one sets up atoms in the \textquotedblleft buffer
gas\textquotedblright\ to be resonant with the probe light (in
this case the \textquotedblleft buffer gas\textquotedblright\ no
longer only acts as a buffer to cool down the EIT atoms ), the
\textquotedblleft buffer gas\textquotedblright\ just plays the
role of a coherent medium; that the photon will be coupled with
collective excitations of the buffer atoms to form
quasi-particles, called polaritons.

Secondly, the study of EIT for photon state storage should be extended to
solid state systems for applications in scalable quantum computing. Here,
EIT atoms with $\Lambda$-type transitions may be realized using solid state
devices, such as artificial atoms based on quantum dots, which are usually
embedded in a solid state substrate. To make such solid state devices as
coherent storage units based on the EIT mechanism, one should consider the
EIT process in the medium of the substrate.

In our study, we first model the medium as a collection of $N$ two-level
atoms, weakly coupled to the quantized probe field \cite{Dutra}. The
\textquotedblleft weak\textquotedblright\ interaction between the atoms and
the probe field is assumed to excite a few atoms, such that the collective
excitations of the atoms behave as bosons. In turn, the photons of the probe
field are dressed by the collective excitations, forming polaritons \cite%
{hopfield} of two modes. According to Hopfield's original paper on quantum
polariton \cite{hopfield} and also according to others \cite{Dutra}\ , such
polariton can be regarded as a macroscopically averaged electromagnetic
field or a displacement field.

We then show that, when one of the two polariton modes is resonant with the
three-level $\Lambda$-type atoms, there still exists a dark state which
decouples from the upper energy level of the $\Lambda $-atom. Utilizing the
dark state, we can adiabatically manipulate the quantum state of the photon
such that the photon state is coherently transferred to the atomic
collective excitation state. We further calculate the susceptibility of the
light propagation in the EIT atomic ensemble embedded in a medium.

In usual case, due to inhomogeneous broadening, ground state
decoherence, loss, etc, the broaden energy levels of atoms in the
EIT ensemble can behave as energy bands and thus limit the
transparency due to the off-resonance of some atoms. Here, the
coherent processes induced by the two-level lossless medium can
only results in a frequency split of effective light field, which
also causes the off-resonance with respect to the fixed energy
levels of EIT atoms. However, by making use of the Hopfield model
\cite{hopfield, Dutra}, the split can be estimated quantitatively
and then one can restore the resonance for the EIT by the
effective light filed in the medium.

The remaining part of this paper is organized as follows. In Sec.~II, we
study the coupled system of quantum light plus medium atoms, and describe
this using two-mode polaritons. Section~II is devoted to study the EIT
effect of a single three-level atom induced by polaritons. In Sec.~IV, for
an ensemble of three-level atoms, we construct the many-atom dark states
based on the spectra-generating algebra method. The influence of the medium
on quantum state transfer is discussed in Sec.~V. In Sec.~VI, we study the
propagation of dressed light. Finally, conclusions are presented in Sec. VII.

\begin{figure}[tbp]
\includegraphics[bb=0 0  611 800, width=8 cm, clip]{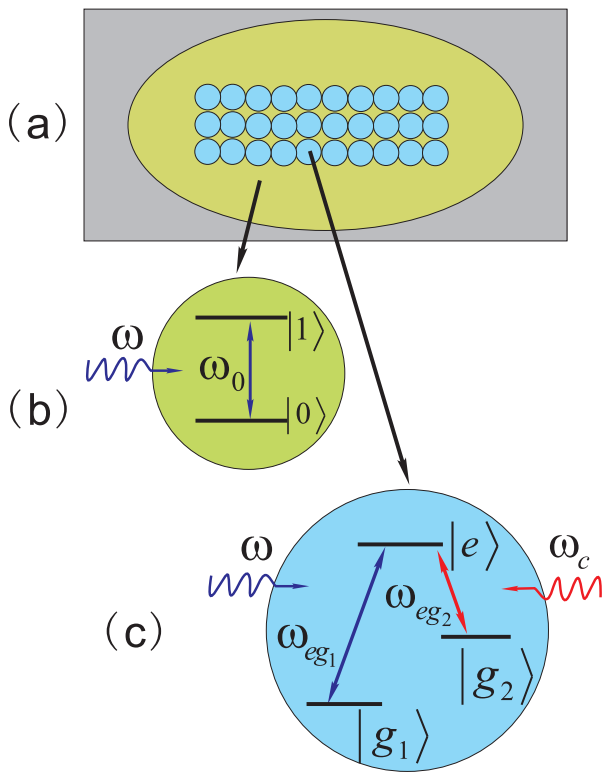}
\caption{(Color online) (a) Schematic diagram of the system under
consideration. (b) The medium [yellow background in (a)] is modeled by $N$
two-level atoms, each medium atom has an identical transition frequency $%
\protect\omega _{0}$. (c) The level structure of the three-level $\Lambda$%
-type atoms.}
\label{media_system}
\end{figure}

\section{Microscopic Hopfield model for material media interacting with a
single-mode cavity field}

\label{sect2}

The system under consideration, shown schematically in Fig.~\ref%
{media_system}, includes a single-mode cavity, a lossless medium, and $M$
identical three-level $\Lambda$-type atoms. The medium is modeled by $N$
two-level atoms with equal level spacing $\omega_0$, and for the $j$th
medium atom, the ground and excited states are denoted, respectively, as $%
|0\rangle_j$ and $|1\rangle_j$. The three-level $\Lambda$-type atoms have
two lower states $|g_1\rangle$ and $|g_2\rangle$ plus an upper state $%
|e\rangle$. A single-mode cavity field $\omega$ is assumed as the probe
field to induce a transition between levels $|e\rangle $ and $|g_{1}\rangle$%
. Finally, a classical control field $\omega_c $ is introduced to couple $%
|e\rangle$ and $|g_{2}\rangle$.

To better understand the effect of the medium, we shall only consider, in
this section, the interaction between the single-mode probe field and the
medium, which is described by the Hamiltonian
\begin{align}
H_{\text{L-M}}& =\hbar \omega a^{\dag }a+\hbar \omega
_{0}\sum_{j=1}^{N}\left\vert 1\right\rangle _{j}\left\langle 1\right\vert
\notag \\
& +\hbar \sum_{j=1}^{N}g_{j}\left( a+a^{\dag }\right) \left( \sigma
_{+}^{\left( j\right) }+\sigma _{-}^{\left( j\right) }\right) ,
\label{hsori}
\end{align}%
where $a^{\dag }$ ($a$) is the creation (annihilation) operator for the
quantum probe field, $\sigma _{+}^{\left( j\right) }=\left\vert
1\right\rangle _{j}\left\langle 0\right\vert $, $\sigma _{-}^{\left(
j\right) }=\left\vert 0\right\rangle _{j}\left\langle 1\right\vert $, and $%
\sigma _{z}^{\left( j\right) }=\left\vert 1\right\rangle _{j}\left\langle
1\right\vert -\left\vert 0\right\rangle _{j}\left\langle 0\right\vert $ are
the quasi-spin Pauli operators for the $j$th atom. Here $g_{j}$ is the
electric-dipole coupling strength between the probe field and the $j$th
atom. For simplicity, we shall assume throughout this paper that $%
g_{j}\equiv g_{\text{medium}}$ is independent of individual atom.

To simplify Hamiltonian (\ref{hsori}), we define the collective quasi-spin
wave operators as
\begin{align*}
B_{k}^{\dagger }& =\frac{1}{\sqrt{N}}\sum_{j=1}^{N}\sigma _{+}^{\left(
j\right) }\exp \left( \frac{2\pi ikj}{N}\right) , \\
B_{k}& =\frac{1}{\sqrt{N}}\sum_{j=1}^{N}\sigma _{-}^{\left( j\right) }\exp
\left( -\frac{2\pi ikj}{N}\right) ,
\end{align*}%
where $k=0,\,\cdots ,\,N-1$. In the large--$N$ limit with low excitation
condition, it was proven that the above collective quasi-spin wave operators
$B_{k}$ and $B_{k^{\prime }}^{\dagger }$ satisfy the bosonic commutation
relations~\cite{Sun,Jin}
\begin{equation}
\left[ B_{k},B_{k^{\prime }}^{\dagger }\right] =\delta _{kk^{\prime }}\,,
\label{col}
\end{equation}%
and
\begin{equation*}
\sum_{j=1}^{N}\,\left\vert 1\right\rangle _{j}\left\langle 1\right\vert
=\sum_{k}B_{k}^{\dag }\,B_{k}.
\end{equation*}%
The commutation relation in Eq.~(\ref{col}) suggests that the low-excitation
behavior of the medium can be described by $N$ bosonic operators, i.e.,
exciton operators. The low energy part of Eq. (\ref{hsori}) then reduces to
Hopfield's Hamiltonian~\cite{hopfield}
\begin{equation}
H_{\text{L-M}}=\hbar \omega a^{\dag }a+\hbar \omega _{0}B_{0}^{\dag
}B_{0}+\hbar G\left( a+a^{\dag }\right) \left( B_{0}+B_{0}^{\dag }\right) ,
\label{hs}
\end{equation}%
where $G=g_{\text{medium}}\,\sqrt{N}\,\propto \sqrt{N/V}$, with $V$ being
the effective volume of the probe field, has a finite van Hove limit. We
remark that, to obtain Eq. (\ref{hs}), we have neglected $N-1$ free exciton
modes $B_{1},\,B_{2},\,\cdots ,\,B_{N-1}$, as they are decoupled from the
probe field.

Equation~(\ref{hs}) can be solved using polariton operators employed by
several authors~\cite{hopfield,Dutra}. Following the procedure given in Ref.~%
\cite{Dutra}, we define the polariton operators as
\begin{equation}
c_{k}=x_{1}^{k}a+y_{1}^{k}a^{\dag }+x_{2}^{k}B_{0}+y_{2}^{k}B_{0}^{\dag }
\label{p_op}
\end{equation}%
with $k=1,\,2$. $c_{k}$ and $c_{k^{\prime }}^{\dag }$ satisfy the usual
bosonic commutation relation $\left[ c_{k},c_{k^{\prime }}^{\dag }\right]
=\delta _{k,k^{\prime }}$ and $\left[ c_{k},c_{k^{\prime }}\right] =0$.
Assuming that the Hamiltonian Eq. (\ref{hs}) is diagonalized by $c_{1}$ and $%
c_{2}$, namely,
\begin{equation*}
H_{\text{L-M}}=\hbar \Omega _{1}c_{1}^{\dag }c_{1}+\hbar \Omega
_{2}c_{2}^{\dag }c_{2},
\end{equation*}%
the coefficients $x_{l}^{k}$ and $y_{l}^{k}$ $(l,\,k=1,\,2)$ are then
obtained via equations
\begin{equation*}
\left[ c_{k},H_{\text{L-M}}\right] \,=\,\hbar \,\Omega \,c_{k}.
\end{equation*}%
Explicitly, we have
\begin{equation*}
x_{j}^{k}=\frac{1}{2}\left( v_{j}^{k}+u_{j}^{k}\right) ,\,y_{j}^{k}=\frac{1}{%
2}\left( v_{j}^{k}-u_{j}^{k}\right) ,\,\,
\end{equation*}%
where
\begin{align*}
u_{1}^{k}& =\frac{\omega }{\Omega _{k}}v_{1}^{k},\,\,\,\,\,\,\,\,\,\,\,\,\,%
\,\,u_{2}^{k}=\frac{\Omega _{k}^{2}-\omega ^{2}}{2G\Omega _{k}}v_{2}^{k}, \\
\,v_{2}^{k}& =\frac{\Omega _{k}^{2}-\omega ^{2}}{2G\omega _{0}}%
v_{1}^{k},\,v_{1}^{k}=\sqrt{\frac{4G^{2}\Omega _{k}\omega _{0}}{\left(
\Omega _{k}^{2}-\omega ^{2}\right) ^{2}+4\omega _{0}\omega G^{2}}}\,,
\end{align*}%
and the eigenfrequencies
\begin{equation}
\Omega _{k}^{2}=\frac{1}{2}\left\{ \omega _{0}^{2}+\omega ^{2}+\left(
-1\right) ^{k}\sqrt{\left( \omega _{0}^{2}-\omega ^{2}\right) ^{2}+16\omega
\,\omega _{0}\,G^{2}}\right\} .  \label{bg_omega}
\end{equation}%
The above results are identical to those obtained using the Hopfield's
approach, as shown in Ref.~\cite{hopfield}, where the effect of the medium
is phenomenologically modeled by many harmonic oscillators. Our results then
indicate that those phenomenological harmonic oscillators essentially
originate from the low-energy collective excitations of the medium atoms. As
a matter of fact, same as the previous treatments for the effect of the
medium ~\cite{Dutra}, our approach also relies on the weak-coupling
assumption, which suggests that the medium effect can be equivalently
studied according to either the two-level model or the harmonic oscillators.

\section{Effect of material media on the dark state of a three-level atom}

\begin{figure}[tbp]
\includegraphics[bb=107 320 509 740, width=6 cm, clip]{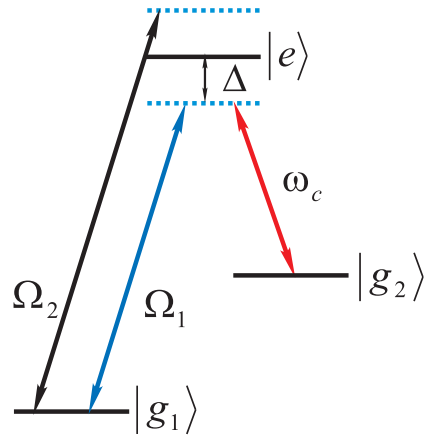}
\caption{(Color online) The coupled system of the two-mode
polaritons and a three-level atom. $\Omega_1$ and $\Omega_2$ are,
respectively, the frequencies of the $c_1$ and $c_2$ modes.
$\protect\omega_c$ is the frequency of the classical control field
and its detuning with the transition frequency
$\protect\omega_{eg_2}$ is $\Delta$. } \label{two_c_EIT}
\end{figure}

In this section, we shall assume that there is only one three-level $\Lambda
$-type atom embedded in the medium. As explained in Sect.~\ref{sect2}, this
atom couples with a single-mode cavity field and a classical light field.
However, due to the effect of the medium, the single cavity mode is
effectively replaced by the two-mode polariton, resulting in the two-color
EIT model shown in Fig.~\ref{two_c_EIT}. The corresponding Hamiltonian takes
the form
\begin{eqnarray}
h &=&\hbar \sum_{i=1}^{2}\Omega _{i}c_{i}^{\dag }c_{i}+\hbar \,\omega
_{eg_{1}}\left\vert e\right\rangle \left\langle e\right\vert +\hbar \left(
\omega _{eg_{1}}-\omega _{eg_{2}}\right) \left\vert g_{2}\right\rangle
\left\langle g_{2}\right\vert  \notag \\
&&+\hbar g\sum_{i=1}^{2}u_{i}\left( c_{i}+c_{i}^{\dag }\right) \left(
\left\vert e\right\rangle \left\langle g_{1}\right\vert +\mathrm{h.c.}\right)
\notag \\
&&+\hbar\, \xi \left( e^{-i\omega _{c}t}\left\vert e\right\rangle
\left\langle g_{2}\right\vert +\mathrm{h.c.}\right) ,  \label{H_single}
\end{eqnarray}%
where $u_{i}=u_{1}^{i}$, $g$ and $\xi $ are the Rabi-frequencies for,
respectively, the single-mode cavity field and the classical light field,
and $\omega _{eg_{1}}$ ($\omega _{eg_{2}}$) is the atomic transition
frequency from the level $\left\vert e\right\rangle $ to the level $%
\left\vert g_{1}\right\rangle $ ($\left\vert g_{2}\right\rangle $) as shown
in Fig.~\ref{media_system}.

We first assume that the classical field and one of two polariton modes, say
$c_{1}$, satisfy the two-photon resonance condition, i.e. $\Omega
_{1}=\omega _{c}+\omega_{eg_{1}}-\omega _{eg_{2}}$. In the interaction
picture and taking the rotating wave approximation, the Hamiltonian Eq. (\ref%
{H_single}) becomes (using $\hbar =1$)
\begin{eqnarray}
h_{I} &=&\widetilde{\Omega }_{2}\,c_{2}^{\dag }c_{2}+\Delta \left\vert
e\right\rangle \left\langle e\right\vert +\xi \left\vert e\right\rangle
\left\langle g_{2}\right\vert  \label{ham_i} \\
&&+\left( \xi \left\vert e\right\rangle \left\langle g_{2}\right\vert
+g\left( u_{1}c_{1}+u_{2}c_{2}\right) \left\vert e\right\rangle \left\langle
g_{1}\right\vert +h.c\right) ,  \notag
\end{eqnarray}%
where $\widetilde{\Omega }_{2}=\Omega _{2}-\omega _{c}-\left( \omega
_{eg_{1}}-\omega _{eg_{2}}\right) $, and $\Delta
=\omega_{eg_1}-\Omega_1=\omega _{eg_{2}}-\omega _{c}$. We note that $h_{I}$
possesses an invariant subspace spanned by the states $\left\vert
e,n_{1},n_{2}\right\rangle $, $\left\vert g_{2},n_{1},n_{2}\right\rangle $, $%
\left\vert g_{1},n_{1}+1,n_{2}\right\rangle $ and $\left\vert
g_{1},n_{1},n_{2}+1\right\rangle $, here $n_{1}$ and $n_{2}$ are the number
of polaritons for modes $c_{1}$ and $c_{2}$, respectively. The matrix
representation of the $h_{I}$ in this invariant subspace is then
\begin{align}
h_{I}& =\left( n_{1}\widetilde{\Omega }_{1}+n_{2}\widetilde{\Omega }%
_{2}\right) I  \notag  \label{eq:7} \\
& +\left(
\begin{array}{cccc}
\Delta & \xi & gu_{1}\sqrt{n_{1}+1} & gu_{2}\sqrt{n_{2}+1} \\
\xi & 0 & 0 & 0 \\
gu_{1}\sqrt{n_{1}+1} & 0 & 0 & 0 \\
gu_{2}\sqrt{n_{2}+1} & 0 & 0 & \widetilde{\Omega }_{2}%
\end{array}%
\right) ,  \notag \\
&
\end{align}%
with $I$ being the identity matrix. Here $h_{I}$ has a zero eigenvalue
corresponding to the eigenstate
\begin{equation}
\left\vert \psi _{0}\right\rangle =\cos \theta \left\vert
g_{1},n_{1}+1,n_{2}\right\rangle -\sin \theta \left\vert
g_{2},n_{1},n_{2}\right\rangle ,  \label{eq:8}
\end{equation}%
where $\theta $ is determined by $\tan \theta =\xi /\left( gu_{1}\sqrt{%
n_{1}+1}\right) $. We immediately notice that $|\psi _{0}\rangle $ is a dark
state formed by polaritons and the two lower atomic levels, in contrast with
that formed directly by photons. Furthermore, $|\psi _{0}\rangle $ can be
factorized as
\begin{eqnarray}
|\psi _{0}\rangle &=&|\phi _{0}\rangle \otimes |n_{2}\rangle  \notag \\
&=&\big(\cos \theta \left\vert g_{1},n_{1}+1\right\rangle -\sin \theta
\left\vert g_{2},n_{1}\right\rangle \big)\otimes |n_{2}\rangle ,
\label{eq:9}
\end{eqnarray}%
where $|\phi _{0}\rangle $ superposes different polariton number states.
Considering now the $n_{1}=0$ subspace, if we manipulate the Rabi-frequency $%
\xi $ of the classical field adiabatically, such that $\theta $ varies from $%
0$ to $\pi /2$, the information of a single polarition state is then stored
into the atomic state.

\section{Collective Atomic Excitation dressed by polaritons}

As the adiabatic manipulation described in the previous section is only
accompanied by a single polariton transfer, it cannot be used to transfer or
store a general state which is a superposition of multiple polariton number
states. To fulfill this purpose, an ensemble of atoms is needed to serve as
the quantum data bus or quantum memory. We, therefore, consider in this
section $M$ identical three-level $\Lambda $-type atoms interacting with the
polariton modes. Same as the single three-level atom case, we assume that
only the $c_1$ mode satisfies the two-photon resonance condition. The
Hamiltonian of the system, in the interaction picture, is (using $\hbar =1$)
\begin{eqnarray}
H_{I}&=&\widetilde{\Omega }_{2}\,c_{2}^{\dag }\,c_{2}+\Delta
\sum_{j=1}^{M}\sigma _{ee}^{\left( j\right) }+\xi
\sum_{j=1}^{M}\sigma_{eg_{2}}^{\left( j\right) }  \notag \\
&&+g\left[\left( u_{1}c_{1}+u_{2}c_{2}\right) \sum_{j=1}^{M}\sigma
_{eg_{1}}^{\left( j\right) }+\mathrm{h.c.}\right],  \label{eq:10}
\end{eqnarray}
where $\sigma _{\mu \nu }^{\left( j\right) }=\left\vert \mu \right\rangle
_{j}\left\langle \nu \right\vert $ ($\mu ,\nu =e,\,g_{1},\,g_{2}$) is a flip
operator of the $j$th atom. To further simplify the notation, we define
collective quasi-spin operators~\cite{Sun}
\begin{eqnarray}
S &=&\sum_{j=1}^{M}\sigma _{ee}^{\left( j\right) },\quad
T_{+}=\sum_{j=1}^{M}\sigma _{eg_{2}}^{\left( j\right) },  \notag \\
\text{\ }A^{\dag } &=&\frac{1}{\sqrt{M}}\sum_{j=1}^{M}\sigma
_{eg_{1}}^{\left( j\right) },  \label{eq:11}
\end{eqnarray}
where $A^\dag$ ($A$) characterizes the collective atomic excitations. We
note that, in the large $M$ limit with low atomic excitations, the operators
$A^{\dag }$ and $A$ satisfy the bosonic commutation relation $\left[
A,\,A^{\dag }\right] =1$. The Hamiltonian Eq.~(\ref{eq:10}) can now be
rewritten as
\begin{eqnarray}  \label{Hi}
H_{I}& =&\widetilde{\Omega }_{2}\,c_{2}^{\dag }\,c_{2}+\Delta S  \notag \\
&& +\left( \xi \,T_{+}+g\,u_{1}\sqrt{M}\,c_{1}A^{\dag }+g\,u_{2}\sqrt{M}%
c_{2}\,A^{\dag }+\mathrm{h.c.}\right) .  \notag \\
\end{eqnarray}%
Following the procedure developed in Ref.~\cite{Sun}, we introduce another
atomic collective excitation operator
\begin{equation}
C=\frac{1}{\sqrt{M}}\sum_{j=1}^{M}\sigma _{g_{1}g_{2}}^{\left( j\right) }.
\label{eq:13}
\end{equation}%
In the large-$M$ and low excitation limit, the collective operators defined
in Eqs.~(\ref{eq:11}) and (\ref{eq:13}) satisfy following basic commutation
relations,
\begin{align}
\left[ A,\,\,\,\,S\right] & =A,\,\,\,\left[ C,\,S\right] =0,\,\left[
A,\,A^{\dag }\right] =1,  \notag \\
\left[ C,\,\,\,C^{\dag }\right] & =1,\,\,\,\left[ T_{+},\,C^{\dag }\right]
=A^{\dag },  \label{commut} \\
\left[ T_{-},\,A^{\dag }\right] & =C^{\dag },\,\,\left[ S,\,\,T_{\pm }\right]
=\pm T_{\pm }.  \notag
\end{align}

%\begin{figure*}[tbp]
%\includegraphics[bb=0 300 540 770, width=6 cm, clip]{adiabatic_control_a.eps}
%\includegraphics[bb=0 300 540 770, width=6 cm, clip]{adiabatic_control_b.eps}
%\caption{Illustration of the influence of medium on the adiabatic
%manipulation. (a) $G=0$. Starting with the state $|X\rangle\equiv\left\vert 0\right\rangle _{C}\otimes \left\vert
%n\right\rangle _{a}\otimes \left\vert 0\right\rangle _{B}$ which encodes the photon information, the system follows the dark state trajectory to state  $|Y\rangle\equiv
%\left\vert n\right\rangle _{C}\otimes \left\vert 0\right\rangle _{a}\otimes
%\left\vert 0\right\rangle _{B}$ during the adiabatic manipulation. This then realizes a perfect quantum state transfer. (b) $G\neq0$. The medium will force the system to deviate from the circle $l$, but the state $X$ will projected into the $X^{\prime }$($%
%\left\vert n\right\rangle _{c_{1}}\otimes \left\vert 0\right\rangle
%_{c_{2}}\otimes \left\vert 0\right\rangle _{C}$). This component will be
%adiabatically controlled to the read state $Y^{\prime }$($\left\vert
%n\right\rangle _{c_{1}}\otimes \left\vert 0\right\rangle _{c_{2}}\otimes
%\left\vert 0\right\rangle _{C}$) in circle $l^{\prime }$, which is also a
%ideal memory state, while its complementary component will not follow this
%circle, so the final state will differ form the ideal memory state $%
%Y^{\prime }$.}
%\label{adiabatic_control}
%\end{figure*}

The effective Hamiltonian Eq.~(\ref{Hi}) is a function of the operators $A$,
$A^{\dag }$, $C$, $C^{\dag }$, and $T_{\pm }$ that generate a close algebra $%
\mathcal{L}$, corresponding to a non-compact group $U\left( \mathcal{L}%
\right) $. This means that the composite system, consisting of photon,
medium, and $\Lambda $-type atoms, possesses a dynamic symmetry of $U\left(
\mathcal{L}\right) $. Using the symmetry analysis~\cite{Sun}, we construct a
dark-state operator of the polariton operator $c_{1}$ and atomic operator $C$
\begin{equation*}
D=c_{1}\cos \theta -C\sin \theta ,
\end{equation*}%
which satisfies $\left[ H_{I},\,D\right] =0$ and $\left[ D,\,D^{\dag }\right]
=1$. Furthermore, we introduce the state
\begin{equation*}
\left\vert \mathbf{0}\right\rangle =\left\vert v\right\rangle \otimes
\left\vert 0\right\rangle _{c_{1}}\otimes \left\vert 0\right\rangle _{c_{2}},
\end{equation*}%
where $\left\vert v\right\rangle =\left\vert g_{1},\,g_{1},\,\cdots
,\,g_{1}\right\rangle $ is the collective ground state with all atoms in
their ground states, $\left\vert 0\right\rangle _{c_{1}}$ and $\left\vert
0\right\rangle _{c_{2}}$ are the vacuum of the polariton modes. We note that
$\left\vert \mathbf{0}\right\rangle $ is an eigenstate of $H_{I}$ with zero
eigenvalue, and consequently, a degenerate class of $H_{I}$ with zero
eigenvalue can be constructed as follows
\begin{equation*}
\left\vert D_{n}\right\rangle =\frac{1}{\sqrt{n!}}D^{\dag n}\left\vert
\mathbf{0}\right\rangle ,
\end{equation*}%
which can be used as a quantum memory. There also exist other eigenstates
with zero eigenvalue; however, as shown in Appendix A, the adiabatic
evolution does not mix them with the dark state $|D_{n}\rangle $.

\section{Quantum adiabatic manipulations in the presence of material media}

In earlier work~\cite{Lukin1,Lukin2,Lukin3,Sun}, the EIT system was proposed
as an efficient quantum memory by adiabatic quantum manipulation. In the
presence of the medium, we explore the possibility of implementing such
quantum manipulation by taking into account the coupling between the quantum
light field and the medium.

The goal of the EIT-based quantum memory is to transmit the information of
the quantum light to the low excitation state of the $\Lambda $-type atomic
ensemble. To see the key point of our studies here, we would like to recall
the basic physical process of the EIT-based quantum storage. If there is no
interaction between the quantum light and the medium ($G=0$), the state used
for quantum storage can be expressed as
\begin{equation*}
\left\vert \Psi \left( \theta \right) \right\rangle =\sum_{n}c_{n}\left\vert
d_{n}\left( \theta \right) \right\rangle \otimes \left\vert 0\right\rangle
_{B},
\end{equation*}%
where $\left\vert 0\right\rangle _{B}$ is the vacuum state of the collective
excitation of the medium atoms and
\begin{equation*}
\left\vert d_{n}\left( \theta \right) \right\rangle =(a^{\dagger }\cos
\theta -C^{\dagger }\sin \theta)^{n}|0\rangle _{C}\otimes |0\rangle _{a}
\end{equation*}%
is a dark state formed by the probe light and the collective excitations of
three-level atoms. Here, $|0\rangle _{C}$ is the vacuum state defined by $C$
and $|0\rangle _{a}$ is the photon vacuum state. Therefore, a perfect
quantum storage of photon states by an atomic ensemble can be realized by
the following adiabatic evolution:
\begin{align*}
\left\vert \Psi \left( \theta =0\right) \right\rangle & =\left\vert
0\right\rangle _{C}\otimes \left( \sum_{n}c_{n}\left\vert n\right\rangle
_{a}\right) \otimes \left\vert 0\right\rangle _{B} \\
\rightarrow \left\vert \Psi \left( \theta =\frac{\pi }{2}\right)
\right\rangle & =\left( \sum_{n}(-1)^nc_{n}\left\vert n\right\rangle
_{C}\right) \otimes \left\vert 0\right\rangle _{a}\otimes \left\vert
0\right\rangle _{B}.
\end{align*}%
%
%
%
%
%
%
%This process is illustrated in Fig.~\ref{adiabatic_control}(a), where the
%adiabatic trajectory of the dark state is represented as a circle
%parameterized by $\theta $.

As shown in the previous section, after we turn on the coupling between the
medium and quantum light, the dark state $\left\vert d_{n}\left( \theta
\right) \right\rangle $ is replaced by $\left\vert D_{n}\left( \theta
\right) \right\rangle $, a dark state formed by the resonant mode $c_{1}$ of
the polariton and the collective excitations of three-level atoms. In this
case, an ideal quantum state transfer should realize the process
\begin{equation*}
\left\vert 0\right\rangle _{C}\otimes \sum_{n}c_{n}\left\vert n\right\rangle
_{a}\otimes \left\vert 0\right\rangle _{B}\rightarrow
\sum_{n}c_{n}\left\vert n\right\rangle _{C}\otimes \left\vert 0\right\rangle
_{c_{1}}\otimes \left\vert 0\right\rangle _{c_{2}}.
\end{equation*}%
However, as we shall show below, quantum state transfer can only be
partially achieved when $G\neq 0$. Without loss of generality, assuming that
the photon state to be transferred is a Fock state, namely, the initial
state of the system is
\begin{equation*}
\left\vert \Psi \left( t=0\right) \right\rangle =\left\vert 0\right\rangle
_{C}\otimes \left\vert n\right\rangle _{a}\otimes \left\vert 0\right\rangle
_{B}.
\end{equation*}%
We note that state $\left\vert n\right\rangle _{a}\otimes \left\vert
0\right\rangle _{B}$ can be expanded using the Fock states of polaritons,
which gives
\begin{eqnarray}
\left\vert \Psi \left( 0\right) \right\rangle &=&S_{n0}\left\vert
0\right\rangle _{C}\otimes \left\vert n\right\rangle _{c_{1}}\otimes
\left\vert 0\right\rangle _{c_{2}}  \notag \\
&&+\sum_{i\neq n\text{ and }j\neq 0}S_{ij}\left\vert 0\right\rangle
_{C}\otimes \left\vert i\right\rangle _{c_{1}}\otimes \left\vert
j\right\rangle _{c_{2}}  \notag \\
&=&S_{n0}\,|D_{n}(\theta =0)\rangle +|\psi ^{\prime }(0)\rangle ,
\label{psi_0}
\end{eqnarray}%
where $|\psi ^{\prime }(0)\rangle =\sum_{i\neq n\text{ and }j\neq
0}S_{ij}\left\vert 0\right\rangle _{C}\otimes \left\vert i\right\rangle
_{c_{1}}\otimes \left\vert j\right\rangle _{c_{2}}$. The coefficients $%
S_{ij} $ can be obtained straightforwardly, in particular, when the coupling
between light and medium atoms is weak, $S_{n0}$ is very close to unity. As
the system evolves adiabatically to time $t$, the wave function becomes
\begin{equation}
\left\vert \Psi \left( t\right) \right\rangle =S_{n0}\,\left\vert
D_{n}\left( \theta (t)\right) \right\rangle +\left\vert \psi ^{\prime
}\left( t\right) \right\rangle ,  \label{d1}
\end{equation}%
and at $\theta (t)=\pi /2$, we have
\begin{equation*}
\left\vert D_{n}\left( \frac{\pi }{2}\right) \right\rangle
=(-1)^{n}\,|n\rangle _{C}\otimes |0\rangle _{c_{1}}\otimes |0\rangle
_{c_{2}}.
\end{equation*}%
Therefore, the first term on the right hand side of Eq. (\ref{d1})
transfers the quanta of the photon to the collective excitation of
the atomic ensemble; the second term, on the other hand, represents
the leakage of the quantum memory.

\begin{figure}[tbp]
\includegraphics[bb=64 221 579  612, width=7
cm, clip]{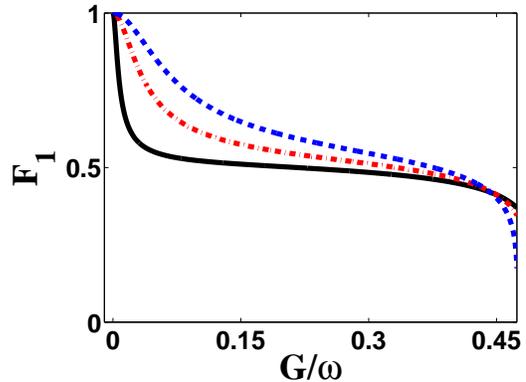} \caption{(Color online) The coupling
strength $G$ dependence of the
one-photon state transmission efficiency $F_1$ for $\protect\omega %
_{0}=0.99\,\protect\omega $ (black solid line), $\protect\omega _{0}=0.95\,%
\protect\omega $ (red dash-dot line), and $\protect\omega _{0}=0.9\,\protect%
\omega $ (blue dashed line).} \label{fidelity}
\end{figure}

%This process is graphically demonstrated in Fig.~\ref{adiabatic_control}(b).

Furthermore, to quantify the effect of the medium, we need to calculate the
efficiency of the $n$-photon state transfer, i.e.,
\begin{equation*}
F_{n}\equiv \left\vert S_{n0}\right\vert ^{2}.
\end{equation*}%
The detailed results on $F_{n}$ are presented in appendix B. In Fig.~\ref%
{fidelity} we plot one-photon state transmission efficiency $F_{1}$ versus
the coupling strength $G$ for different ratio of the frequencies of the
quantum light and the collective excitation of the medium. We see that the
presence of the medium notably reduces the transmission efficiency,
especially when the quantum light is nearly resonant with the collective
excitation of the medium atoms.

We notice that $F_{n}$ actually equal to 1 when $G=0$ in the
resonant case and thus we can not resort to the same calculation
method about the transmission efficiency shown in the appendix B.
When $\omega _{0}=\omega $ together with $G=0$, there would be of
singularity for the transmission efficiency if we carried out the
same calculation as that in the appendix B. Physically, we can
consider the cases with small detuning $\omega _{0}-\omega $ and
there is not an obvious jump of the transmission efficiency as
shown in Fig. 3. Actually, there is a jump of $F_1 $ from 1 to 0.5
in the resonant case when we turn on the coupling G from $0$ to a
small value. The strict resonant condition is never feasible in
the practical experiment, thus we only consider two nearly
resonant cases in Fig. 3.

\section{Propagation of the dressed quantum light}

\begin{figure}[tbp]
\includegraphics[bb=107 320 509 740, width=6 cm, clip]{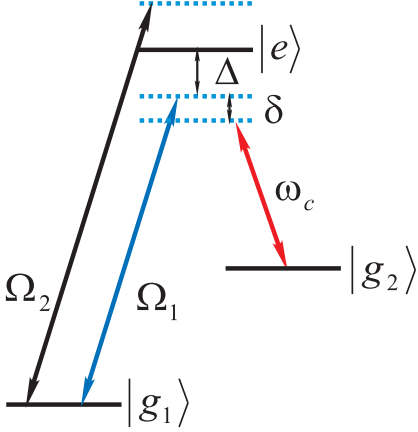}
\caption{(Color online) Same as Fig.~\protect\ref{two_c_EIT}, except that
here the $c_{1}$ polariton mode and the classical light field do not satisfy
the two-photon resonant condition, i.e.\,, $\protect\delta =\Omega _{1}-%
\protect\omega _{c}-\protect\omega _{g_{2}}\neq 0$.}
\label{propagation}
\end{figure}

To consider the dynamical process of a quantum state transfer, which is
usually described by the slowing and the stopping of light, we study the
dispersion and absorption properties of the dressed quantum light,
propagating in a $\Lambda $-type atomic ensemble. To achieve our goal, we
consider the case when the $c_{1}$ mode and the classical light field have a
small two-photon detuning $\delta =\Omega _{1}-\omega _{c}-\omega
_{g_{2}}\ll \Omega _{1}$ (see Fig.~\ref{propagation}). The Hamiltonian in
the interaction picture now becomes
\begin{eqnarray*}
H_{I}&=& \widetilde{\Omega }_{2}c_{2}^{\dag }c_{2}+\Delta S+e^{i\delta t}\xi
T_{+} \\
&&+gu_{1}\sqrt{N}c_{1}A^{\dag }+gu_{2}\sqrt{N}c_{2}A^{\dag }+h.c.~.
\end{eqnarray*}%
With the help of the basic commutation relations in Eq.~(\ref{commut}), we
can approximately write down the Heisenberg equations for the operators $%
A,\,C$, and $c_{2}$,
\begin{eqnarray}
\dot{A}&=& -\Gamma _{A}\,A-i\,\Delta\, A-i\,g\,u_{1}\sqrt{N}\,c_{1}  \notag
\\
&& -ie^{i\delta t}\,\xi C-i\,g\,u_{2}\,\sqrt{N}\,c_{2},  \notag \\
\dot{C}&=& -\Gamma _{C}\,C-ie^{-i\delta t}\,\xi\, A, \\
\dot{c}_{2}&=& -\Gamma _{c_{2}}\,c_{2}-i\widetilde{\Omega }%
_{2}\,c_{2}-i\,g\,u_{2}\,\sqrt{N}A,  \notag
\end{eqnarray}%
where we have ignored quantum fluctuations since we only calculate
the group velocity. In addition, we have phenomenologically
introduced the damping rate $\Gamma_{c_{2}} $ of the mode $c_{2}$
and the decay rates $\Gamma _{A}$ and $\Gamma_{C}$ for,
respectively, the states $\left\vert e\right\rangle $ and
$\left\vert g_{2}\right\rangle $. $\Gamma_A$ and $\Gamma_C$ can be
estimated as the spontaneous emission rates of the respective
levels, which are proportional to the cube of the atomic
transition frequencies, we therefore have $\Gamma _{A}\gg\Gamma
_{C}$. The damping rate of the polariton mode is mainly due to the
spontaneous emissions of the medium atoms and the leakage of the
cavity.  Since the former is negligible for a lossless medium, we
can further assume $\Gamma_{C}\gg \Gamma _{c_{2}}$ for a high
quality cavity.

\begin{figure}[ptb]
\includegraphics[bb=50 228 410 588, width=8.5cm, clip]{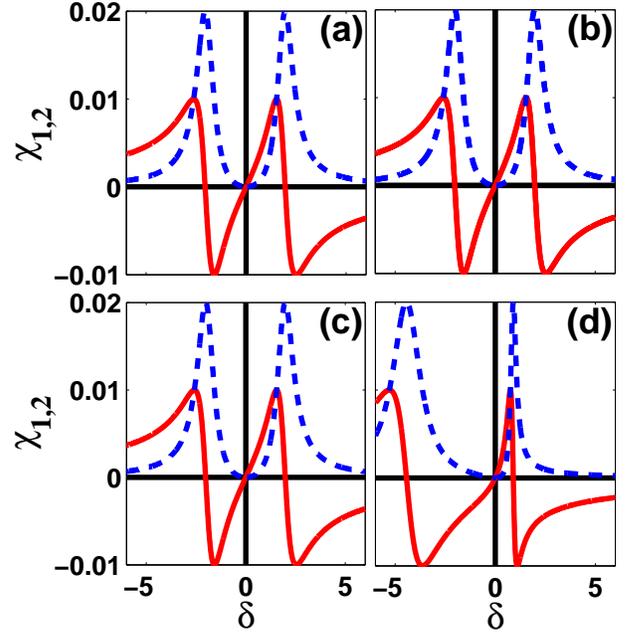}
\caption{(Color online) $\protect\chi_{1}$ (red solid lines) and $\protect%
\chi_{2}$ (blue dashed lines) versus the two-photon detuning $\protect\delta$
(in units of $\Gamma_A $, i.e., here $\Gamma_A=1$). (a) $\protect\omega %
_{0}=0.9\,\protect\omega$ and $G=0$; (b) $\protect\omega_{0}=0.9\,\protect%
\omega$ and $G=0.1\,\protect\omega$; (c) $\protect\omega _{0}=\protect\omega$
and $G=0.1\,\protect\omega$; (d) $\protect\omega_{0}=\protect\omega$ and $%
G=0.001\,\protect\omega$. Other parameters are $\Gamma_{A}=1$, $%
\Gamma_{C}=10^{-4}$, $\Gamma_{c_{2}}=10^{-6},g \protect\sqrt{N}=100$, and $%
\protect\omega=10^6$.}
\label{permittivity}
\end{figure}

To find a steady state solution for the above equations of motion, it is
convenient to remove the fast-oscillating factors, by making the
transformation $C=e^{-i\delta t}C^{\prime }$, which yields
\begin{eqnarray}
\dot{A}& =&-\Gamma _{A}\,A-i\,\Delta \,A-i\,g\,u_{1}\sqrt{N}c_{1}  \notag \\
&&-i\xi C^{\prime }-igu_{2}\sqrt{N}\,c_{2},  \notag \\
\dot{C}^{\prime }& =&-\Gamma _{C}\,C^{\prime }-i\xi A+i\,\delta\,C^{\prime },
\\
\dot{c}_{2}& =&-\Gamma _{c_{2}}\,c_{2}-i\widetilde{\Omega }%
_{2}\,c_{2}-i\,g\,u_{2}\sqrt{N}A.  \notag
\end{eqnarray}%
The steady state solution can be obtained by letting $\dot{A}=\dot{C}%
^{\prime }=\dot{c}_{2}=0$, from which we find the mean value of $A$ as
\begin{equation}
\left\langle A\right\rangle =\frac{-igu_{1}\alpha \beta \left\langle
c_{1}\right\rangle }{\alpha \beta \left( \Gamma _{A}+i\Delta \right) +\beta
\xi ^{2}+\alpha g^{2}u_{2}^{2}N}  \label{A}
\end{equation}%
with $\alpha =\Gamma _{C}-i\delta $ and $\beta =\Gamma _{c_{2}}+i\widetilde{%
\Omega }_{2}$.

It is noted here that the dressed quantum light or the propagating
polaritons nearly on resonance with the $\Lambda $-type atom can be
described by
\begin{align*}
E\left( t\right) & =\varepsilon e^{-i\Omega _{1}t}+h.c. \\
& =u_{1}\sqrt{\frac{\omega }{2V\epsilon _{0}}}c_{1}e^{-i\Omega _{1}t}+h.c.,
\end{align*}%
where $\epsilon _{0}$ is the permittivity of free space and $V$ is the
effective mode volume, which, for simplicity, is chosen to be equal to the
interaction volume. In this case, the time-independent part of the polariton
field strength is $u_{1}\sqrt{\omega /2V\epsilon _{0}}.$ We remark here that
the Hopfield polariton field can be understood as a displacement field or
macroscopic electromagnetic field corresponding to the polarization
\begin{equation}
\left\langle P\right\rangle =\left\langle p\right\rangle e^{-i\Omega _{1}t}+%
\mathrm{h.c.}\,=\,\epsilon _{0}\,\chi \left\langle \varepsilon \right\rangle
e^{-i\Omega _{1}t}+\mathrm{h.c.}\,,  \label{P}
\end{equation}%
where $\chi $ is the susceptibility. After neglecting the effect of the
non-resonance polariton, the average polarization
\begin{equation}
\left\langle p\right\rangle =\frac{\mu }{V}\left\langle \sum_{j=1}^{N}\sigma
_{eg_{2}}^{\left( j\right) }\right\rangle =\frac{\mu \sqrt{N}}{V}%
\left\langle A\right\rangle  \label{p}
\end{equation}%
can also be expressed in terms of the average of the exciton operator $A.$
Combining the Eqs.~(\ref{A}-\ref{p}), we obtain %\begin{widetext}%
\begin{equation}
\chi =\frac{i2g^{2}N\alpha \beta }{\omega \left[ \alpha \,\beta \left(
\Gamma _{A}+i\Delta \right) +\beta \,\xi ^{2}+\alpha \,g^{2}u_{2}^{2}N\right]
}.  \label{bg_x}
\end{equation}%
%
%
%
%
%
%
%
%
%
%\end{widetext}
The real and imaginary parts $\chi _{1}$ and $\chi _{2}$ of the complex
susceptibility $\chi =\chi _{1}+i\chi _{2}$ can be explicitly expressed as%
\begin{align*}
\chi _{1}& =\frac{\left( \delta \Gamma _{c_{2}}-\Gamma _{C}\widetilde{\Omega
}_{2}\right) \Theta +\left( \Gamma _{C}\Gamma _{c_{2}}+\widetilde{\Omega }%
_{2}\delta \right) \Xi }{\Theta ^{2}+\Xi ^{2}}F, \\
\chi _{2}& =\frac{\left( \Gamma _{C}\Gamma _{c_{2}}+\widetilde{\Omega }%
_{2}\delta \right) \Theta -\left( \delta \Gamma _{c_{2}}-\Gamma _{C}%
\widetilde{\Omega }_{2}\right) \Xi }{\Theta ^{2}+\Xi ^{2}}F,
\end{align*}%
where $F=2g^{2}N/\omega $, and
\begin{eqnarray*}
\Theta & =&\Gamma _{C}\left( \Gamma _{A}\Gamma _{c_{2}}-\Delta \widetilde{%
\Omega }_{2}\right) +\xi ^{2}\Gamma _{c_{2}} \\
&& +\delta \left( \Delta \Gamma _{c_{2}}+\widetilde{\Omega }_{2}\Gamma
_{A}\right) +g^{2}u_{2}^{2}N\Gamma _{C}, \\
\Xi & =&-\delta \left( \Gamma _{A}\Gamma _{c_{2}}-\Delta \widetilde{\Omega }%
_{2}\right) +\widetilde{\Omega }_{2}\xi ^{2} \\
&& +\Gamma _{C}\left( \Delta \Gamma _{c_{2}}+\widetilde{\Omega }_{2}\Gamma
_{A}\right) -\delta g^{2}u_{2}^{2}N.
\end{eqnarray*}%
It is well known that $\chi _{1}$ and $\chi _{2}$ are related to the
dispersion and absorption, respectively. In Fig.~\ref{permittivity}, $\chi
_{1}$ and $\chi _{2}$ are plotted versus the two-photon detuning $\delta $.

Figure.~\ref{permittivity}(a) shows the case (i) where there is no coupling
between the quantum light and the medium. The result is obviously the same
as that of the conventional EIT effects. Figure.~\ref{permittivity}(b) and
Fig.~\ref{permittivity}(c) demonstrate almost the same dispersion and
absorption properties as that in the case without the influence of medium
shown in Fig.~\ref{permittivity}(a). Figure~\ref{permittivity}(b) describes
the case (ii) that the frequency of the quantum light $\omega$ and the
collective excitation frequency of the medium $\omega_{0}$ is largely
detuned in comparison with the parameters $g^{2}u_{2}^{2}N,\delta$ and $%
\Delta$, i.e.,%
\begin{equation*}
\left\vert \omega-\omega_{0}\right\vert \gg\delta,\,\Delta,\,g^{2}u_{2}^{2}N.
\end{equation*}
Figure.~\ref{permittivity}(c) describes the case (iii) that the coupling
strength $G$ between the quantum light medium is larger than the parameters $%
g^{2}u_{2}^{2}N,\,\delta$ and $\Delta$, i.e., $G\,\gg\,\delta,\,\Delta,%
\,g^{2}u_{2}^{2}N$.

The above phenomenon, predicted by our numerical calculations, can be well
explained. In the configuration, illustrated in Fig.~\ref{propagation}, when
the mode $c_{1}$ of the polariton is nearly resonant with respect to the
transition between $\left\vert e\right\rangle $ and $\left\vert
g_{1}\right\rangle $, the role of the mode $c_{2}$ can be neglected if this
mode is off-resonace with respect to the transitions from $|e\rangle$ to $%
|g_{1}\rangle$ and $|g_{2}\rangle$. Then the system will be reduced to the
conventional EIT model, where the mode $c_{1}$ of the polariton plays the
same role as that of the quantum light, taken as the probe field. This case
must lead to the same result as the conventional EIT case about the
dispersion and absorption even in the presence of the medium. Now we can
show that both cases (ii) and (iii) can give rise to the condition that the
frequencies of the mode $c_{1}$ and $c_{2}$, i.e. $\Omega_{1}$ and $%
\Omega_{2}$, are largely detuned as mentioned above. We note that
\begin{equation*}
\left\vert \Omega_{2}-\Omega_{1}\right\vert =\left( \Omega_{2}+\Omega
_{1}\right) ^{-1}\sqrt{\left( \omega_{0}^{2}-\omega^{2}\right)
^{2}+16\omega\omega_{0}G^{2}}.
\end{equation*}
In the case (ii) with $\left\vert \omega-\omega_{0}\right\vert \gg
\delta,\,\Delta,\,g^{2}u_{2}^{2}N$, combining the condition
\begin{equation*}
\sqrt{\left( \omega_{0}^{2}-\omega^{2}\right) ^{2}+16\omega\omega_{0}G^{2}}%
>\left\vert \omega_{0}^{2}-\omega^{2}\right\vert
\end{equation*}
with the condition $\Omega_{2}+\Omega_{1}$ being of the order of $%
\omega+\omega_{0}$, we can find that $\left\vert \Omega_{2}-\Omega
_{1}\right\vert $ is in the order of $\left\vert
\omega-\omega_{0}\right\vert $. This implies that $\left\vert
\Omega_{2}-\Omega_{1}\right\vert \gg \delta,\,\Delta,\,g^{2}u_{2}^{2}N$;
which shows that the large-detuning condition is satisfied. The same
analysis can also be applied to the case (iii) if we note the condition
\begin{equation*}
\sqrt{\left( \omega_{0}^{2}-\omega^{2}\right) ^{2}+16\omega\,\omega
_{0}\,G^{2}}>4\sqrt{\omega\omega_{0}}G.
\end{equation*}
Due to $|\widetilde{\Omega}_{2}|\approx\left\vert \Omega_{2}-\Omega
_{1}\right\vert $, if%
\begin{equation*}
\left\vert \Omega_{2}-\Omega_{1}\right\vert
\gg\delta,\,\Delta,\,g^{2}u_{2}^{2}N,
\end{equation*}
we can neglect all the terms, which do not have a factor of $\widetilde {%
\Omega}_{2}$ in the denominator and numerator of Eq.~(\ref{bg_x}). After
calculations, we can obtain the same expression of the susceptibility as
that in the conventional EIT case~\cite{Li}.

For the result illustrated in Fig.~\ref{permittivity}(d), where the
parameters are assumed to satisfy the condition $\left\vert
\omega-\omega_{0}\right\vert /\omega\ll1$ and $G/\omega\ll1$, we can obtain
the results, about the dispersion and the absorption, which are different
from those in the conventional EIT. The phenomenon is the deformed
transparent window which is assisted by the collective excitation of the
medium. Indeed, if $\left\vert \omega-\omega_{0}\right\vert /\omega\ll1$ and
$G/\omega\ll1, $ $|\widetilde{\Omega}_{2}|\approx\left\vert
\Omega_{2}-\Omega_{1}\right\vert $ is smaller than or of the same order of $%
\delta$ and $\Delta.$ So the term related to the mode $c_{2} $ of the
polariton, i.e.\,, $g^{2}u_{2}^{2}N\left( \Gamma_{C}-i\delta\right) $, has a
dominant effect on the susceptibility $\chi.$

\section{Conclusions}

In conclusion, we have studied the influence of a lossless medium on an EIT
system. We find that even in the presence of the medium, the whole system
still has dark states. This implies that, in some cases, the EIT system in
the medium can still serve as a quantum memory. We also calculate the
dispersion and absorption properties of the dressed quantum light. We find
that the result obtained here is quite different from that of the
conventional EIT. If the coupling strength between the quantum light and the
medium is sufficiently strong, the ensemble of the three-level atoms with $%
\Lambda$-type transitions can easily become transparent, if the usual EIT
approach is applied.

\section{Acknowledgments}

CPS is supported by the NSFC with grant No. 90203018, 10474104 and 60433050,
and NFRPC with No. 2001CB309310 and 2005CB724508. SY is supported by NSFC
No. 10674141. FN was supported in part by the National Security Agency
(NSA), Laboratory of Physical Sciences (LPS), and Army Research Office
(ARO); and by the National Science Foundation grant No.~EIA-0130383.

\appendix

\section{Dynamic symmetry analysis of the system}

\label{app:appendix a}

Starting from the dark state $\left\vert D_{n}\right\rangle $, we can use
the spectrum--generating algebra method \cite{algebra} to build other
eigenstates of the whole system. We now introduce the bright-state polariton
operator
\begin{equation*}
B=c_{1}\sin\theta+C\cos\theta,
\end{equation*}
which satisfies
\begin{equation*}
\left[ B,B^{\dag}\right] =1,\,\left[ B,D^{\dag}\right] =0,\,\left[ B,D\right]
=0.
\end{equation*}

It is straightforward to obtain the commutation relations
\begin{align*}
\left[ H_{I},B^{\dag }\right] & =\varepsilon B^{\dag }, \\
\left[ H_{I},A^{\dag }\right] & =\Delta A^{\dag }+\varepsilon B^{\dag
}+gu_{2}\sqrt{N}c_{2}^{\dag }, \\
\left[ H_{I},c_{2}^{\dag }\right] & =\widetilde{\Omega }_{2}c_{2}^{\dag
}+gu_{2}\sqrt{N}A^{\dag },
\end{align*}%
with $\varepsilon =\sqrt{g^{2}u_{1}^{2}N\xi ^{2}}$. We can introduce three
independent bosonic operators
\begin{equation*}
Q_{i}=\eta _{1}^{i}A+\eta _{2}^{i}B+\eta _{3}^{i}C,\,i=1,\,2,\,3,
\end{equation*}%
which satisfy%
\begin{equation*}
\left[ Q_{i},Q_{j}^{\dag }\right] =\delta _{ij},\,\left[ Q_{i},Q_{j}\right]
=0
\end{equation*}%
to diagonalize the Hamiltonian $H_{I}$, i.e.
\begin{equation*}
\left[ Q_{i},H_{I}\right] =\epsilon _{i}Q_{i}.
\end{equation*}

Based on the above commutation relations, we can construct eigenstates
\begin{equation*}
\left\vert e\left( m_{1},m_{2},m_{3},n\right) \right\rangle =\frac{%
Q_{1}^{\dag m_{1}}Q_{2}^{\dag m_{2}}Q_{3}^{\dag m_{3}}}{\sqrt{%
m_{1}!m_{2}!m_{3}!}}\left\vert D_{n}\right\rangle ,
\end{equation*}%
of the whole system, corresponding to eigenvalues
\begin{equation*}
E\equiv E\left( m_{1},m_{2},m_{3}\right) =m_{1}\epsilon _{1}+m_{2}\epsilon
_{2}+m_{3}\epsilon _{3},
\end{equation*}%
with $m_{1},\,m_{2},\,m_{3}=0,\,1,\,2,\cdots $. The above equations show
that there exists a larger class of states %\begin{widetext}%
\begin{equation*}
\mathbf{S}:\left\{ \left\vert e\left( \{m_{i}\},n\right) \right\rangle
\equiv \left\vert D\left( \{m_{i}\},n\right) \right\rangle |n=0,\,1,\,\cdots
;E=0\right\}
\end{equation*}

%
%
%
%
%\end{widetext}
with zero eigenvalue $E\equiv E\left( m_{1},m_{2},m_{3}\right) =0$, here, $%
\{m_{i}\}\equiv m_{1},m_{2},m_{3}$. But we can show bellow that these states
with zero eigenvalue do not mix with each other under adiabatic
manipulation. Any state
\begin{equation*}
\left\vert \phi \left( t\right) \right\rangle
=\sum_{m_{1}m_{2}m_{3}n}c_{m_{1}m_{2}m_{3}n}\left\vert D\left(
m_{1},m_{2},m_{3},n\right) \right\rangle ,
\end{equation*}%
with zero eigenvalue evolves according to
\begin{equation*}
i\frac{d}{dt}c_{m_{1}m_{2}m_{3}n}\left( t\right) =\sum_{m_{1}^{\prime
}m_{2}^{\prime }m_{3}^{\prime }n^{\prime
}}D_{m_{1}m_{2}m_{3}n}^{m_{1}^{\prime }m_{2}^{\prime }m_{3}^{\prime
}n^{\prime }}c_{m_{1}^{^{\prime }}m_{2}^{^{\prime }}m_{3}^{^{\prime
}}n^{^{\prime }}}+F,
\end{equation*}%
where $F$ is a certain functional of the eigenstates with non-zero
eigenvalues, which can be neglected under the adiabatic conditions~\cite%
{Sun-prd,Zee}, and%
\begin{equation*}
iD_{m_{1}m_{2}m_{3}n}^{m_{1}^{\prime }m_{2}^{\prime }m_{3}^{\prime
}n^{\prime }}=\left\langle D\left( m_{1}^{\prime },m_{2}^{\prime
},m_{3}^{\prime },n\right) \left\vert \partial _{t}\right\vert D\left(
m_{1},m_{2},m_{3},n\right) \right\rangle .
\end{equation*}%
We note that $\partial _{\theta }B=D$ and $\partial _{\theta }D=B$, and we
have
\begin{equation*}
\partial _{t}\left\vert D\left( m,n\right) \right\rangle =\dot{\theta}%
\partial _{\theta }\left\vert D\left( m_{1},m_{2},m_{3},n\right)
\right\rangle ,
\end{equation*}%
where $\partial _{\theta }\left\vert D\left( m_{1},m_{2},m_{3},n\right)
\right\rangle $ includes six terms: $\left\vert e\left( m_{1}\mp
1,m_{2},m_{3},n\pm 1\right) \right\rangle ,\left\vert e\left( m_{1},m_{2}\mp
1,m_{3},n\pm 1\right) \right\rangle $ and $\left\vert e\left(
m_{1},m_{2},m_{3}\mp 1,n\pm 1\right) \right\rangle $ which are all
eigenstates with non-zero eigenvalues. This implies the exact result
\begin{equation*}
\left\langle D\left( m_{1}^{\prime },m_{2}^{\prime },m_{3}^{\prime
},n\right) \left\vert \partial _{t}\right\vert D\left(
m_{1},m_{2},m_{3},n\right) \right\rangle =0,
\end{equation*}%
showing that there is no mixing among the states with zero eigenvalue during
the adiabatic evolution.

\section{Calculation of the transmission efficiency}

\label{app:appendix b}

Starting from the Hamiltonian (\ref{hs}), we now calculate the transmission
efficiency in the coordinate representation. We recall the relation between
the operators $a^{\dag },\,a,\,B^{\dag },\,B$ and the corresponding
coordinate operators and moment operators $x_{1},\,x_{2},\,p_{1},\,$and $%
p_{2},$ i.e.%
\begin{align}
x_{1}& =\sqrt{\frac{\hbar }{2m\omega }}\left( a^{\dag }+a\right) \text{,} \\
\text{ }p_{1}& =i\sqrt{\frac{m\hbar \omega }{2}}\left( a^{\dag }-a\right) ,
\\
x_{2}& =\sqrt{\frac{\hbar }{2m\omega }}\left( B^{\dag }+B\right) \text{, } \\
p_{2}& =i\sqrt{\frac{m\hbar \omega }{2}}\left( B^{\dag }-B\right) ,
\end{align}%
where $m$ is the mass of the oscillator. We have here assumed the mass of
the two oscillators to be the same. Therefore, we have the Hamiltonian
\begin{equation}
H_{\text{L-M}}=\frac{1}{2m}\left( p_{1}^{2}+p_{2}^{2}\right) +\frac{1}{2}%
\left( Ax_{1}^{2}+Bx_{2}^{2}+Cx_{1}x_{2}\right)
\end{equation}%
for two coupled harmonic oscillators. Here, we have neglect the zero point
energy, and
\begin{equation}
A=m\omega ^{2}\text{, }B=m\omega _{0}^{2}\text{, }C=4Gm\sqrt{\omega \omega
_{0}}\,\text{. }
\end{equation}

Let $\omega \neq \omega _{0}$ and define the canonical coordinates
\begin{equation}
\left(
\begin{array}{c}
y_{1} \\
y_{2}%
\end{array}%
\right) =\left(
\begin{array}{cc}
\cos \frac{\alpha }{2} & -\sin \frac{\alpha }{2} \\
\sin \frac{\alpha }{2} & \cos \frac{\alpha }{2}%
\end{array}%
\right) \left(
\begin{array}{c}
x_{1} \\
x_{2}%
\end{array}%
\right) ,
\end{equation}%
where

\
\begin{equation}
\tan \alpha =\frac{C}{B-A}.
\end{equation}%
Then we diagonalize $\ H_{\text{L-M}}$ with two decoupled harmonic
oscillators
\begin{equation}
H_{\text{L-M}}=\frac{1}{2m}\left( p_{y_{1}}^{2}+p_{y_{2}}^{2}\right) +\frac{1%
}{2}\left( K_{1}y_{1}^{2}+K_{2}y_{2}^{2}\right)
\end{equation}%
with%
\begin{align}
K_{1}& =\frac{A+B-K}{2},  \notag \\
K_{2}& =\frac{A+B+K}{2},
\end{align}%
and
\begin{equation}
K=\left( B-A\right) \sqrt{1+\frac{C^{2}}{\left( B-A\right) ^{2}}}\,.
\end{equation}%
Therefore, the eigenenergy of the above system is
\begin{equation}
E=\hbar \Omega _{1}\left( n_{1}+\frac{1}{2}\right) +\hbar \Omega _{2}\left(
n_{2}+\frac{1}{2}\right) ,
\end{equation}%
for $\,n_{1},\,\,n_{2}=0,\,1,\,2,\cdots ,\,$and the corresponding eigenstate
$\left\vert n_{1}\right\rangle _{c_{1}}\otimes \left\vert n_{2}\right\rangle
_{c_{2}}$ is expressed as
\begin{eqnarray*}
\psi _{n_{1}n_{2}}\left( y_{1},y_{2}\right) &=&\mathcal{N}_{n_{1}}^{\left(
b_{1}\right) }\mathcal{N}_{n_{2}}^{\left( b_{2}\right) }\exp \left\{ {-\frac{%
1}{2}b_{1}^{2}y_{1}^{2}-\frac{1}{2}b_{2}^{2}y_{2}^{2}}\right\} \\
&&\times H_{n_{1}}\left( b_{1}y_{1}\right) H_{n_{1}}\left( b_{2}y_{2}\right)
,
\end{eqnarray*}%
in terms of the $n$th--order Hermite polynomial $H_{n}\left( \xi \right) $
with
\begin{align}
\mathcal{N}_{n_{1}}^{\left( b_{1}\right) }& =\left[ \frac{b_{1}}{\sqrt{\pi }%
2^{n_{1}}n_{1}!}\right] ^{\frac{1}{2}},  \notag \\
\mathcal{N}_{n_{2}}^{\left( b_{2}\right) }& =\left[ \frac{b_{2}}{\sqrt{\pi }%
2^{n_{2}}n_{2}!}\right] ^{\frac{1}{2}}, \\
b_{1}& =\left( \frac{mK_{1}}{\hbar }\right) ^{1/4},  \notag \\
b_{2}& =\left( \frac{mK_{2}}{\hbar }\right) ^{1/4}.  \notag
\end{align}%
When there is no coupling between the medium and the quantum light, the
Hamiltonian of the medium and quantum light is
\begin{equation}
H_{\text{uncoupled}}=\frac{1}{2m}\left( p_{1}^{2}+p_{2}^{2}\right) +\frac{1}{%
2}\left( Ax_{1}^{2}+Bx_{2}^{2}\right) .
\end{equation}%
Therefore the eigenenergies and eigenstates $\left\vert n_{1}\right\rangle
_{a}\otimes \left\vert n_{2}\right\rangle _{B}$ are given by
\begin{equation}
E=\hbar \omega \left( n_{1}+\frac{1}{2}\right) +\hbar \omega _{0}\left(
n_{2}+\frac{1}{2}\right) ,\,
\end{equation}%
and
\begin{eqnarray*}
\phi _{n_{1}n_{2}}\left( x_{1},x_{2}\right) &=&\mathcal{N}_{n_{1}}^{\left(
a_{1}\right) }\mathcal{N}_{n_{2}}^{\left( a_{2}\right) }\exp \left\{ -\frac{1%
}{2}a_{1}^{2}x_{1}^{2}-\frac{1}{2}a_{2}^{2}x_{2}^{2}\right\} \\
&&\times H_{n_{1}}\left( a_{1}x_{1}\right) H_{n_{1}}\left( a_{2}x_{2}\right)
\end{eqnarray*}%
respectively, for $n_{1},\,\,n_{2}=0,\,1,\,2,\cdots \,\,$and%
\begin{align*}
\mathcal{N}_{n_{1}}^{\left( a_{1}\right) }& =\left[ \frac{a_{1}}{\sqrt{\pi }%
2^{n_{1}}n_{1}!}\right] ^{\frac{1}{2}}, \\
\mathcal{N}_{n_{2}}^{\left( a_{2}\right) }& =\left[ \frac{a_{2}}{\sqrt{\pi }%
2^{n_{2}}n_{2}!}\right] ^{\frac{1}{2}}, \\
a_{1}& =\left( \frac{mA}{\hbar }\right) ^{1/4}, \\
a_{2}& =\left( \frac{mB}{\hbar }\right) ^{1/4}.
\end{align*}

Now we can calculate the transmission efficiency of $n$-photon state%
\begin{eqnarray}
F_{n} &=&\left\vert S_{n0}\right\vert ^{2}  \notag \\
&=&\left\vert \left. _{c_{2}}\left\langle 0\right\vert \right. \otimes
\left. _{c_{1}}\left\langle n\right\vert \otimes \left\langle v\right\vert
\right. \left\vert v\right\rangle \otimes \left\vert n\right\rangle
_{a}\otimes \left\vert 0\right\rangle _{B}\right\vert ^{2}  \notag \\
&=&\left\vert \int \int \psi _{n0}\left[ y_{1}\left( x_{1},x_{2}\right)
,y_{2}\left( x_{1},x_{2}\right) \right] \phi _{n0}\left[ x_{1},x_{2}\right]
dx_{1}dx_{2}\right\vert ^{2}  \notag \\
&=&\left\vert \int \int \exp \left\{ -\frac{1}{2}\left(
a_{1}^{2}x_{1}^{2}+a_{2}^{2}x_{2}^{2}+b_{1}^{2}y_{1}^{2}+b_{2}^{2}y_{2}^{2}%
\right) \right\} \right. \\
&&\times H_{n}\left( a_{1}x_{1}\right) H_{0}\left( a_{2}x_{2}\right)
H_{n}\left( b_{1}y_{1}\right) H_{0}\left( b_{2}y_{2}\right)  \notag \\
&&\left. \times \mathcal{N}_{n}^{\left( b_{1}\right) }\mathcal{N}%
_{0}^{\left( b_{2}\right) }\mathcal{N}_{n}^{\left( a_{1}\right) }\mathcal{N}%
_{0}^{\left( a_{2}\right) }dx_{1}dx_{2}\right\vert ^{2}\,.  \notag
\end{eqnarray}%
To explicitly calculate the integral in the above formula we can define a
new pair of coordinates,
\begin{equation}
\left(
\begin{array}{c}
z_{1} \\
z_{2}%
\end{array}%
\right) =\left(
\begin{array}{cc}
\cos \frac{\beta }{2} & -\sin \frac{\beta }{2} \\
\sin \frac{\beta }{2} & \cos \frac{\beta }{2}%
\end{array}%
\right) \left(
\begin{array}{c}
x_{1} \\
x_{2}%
\end{array}%
\right) \text{, }
\end{equation}%
where%
\begin{equation*}
\tan \beta =\frac{R}{Q-P},
\end{equation*}%
\begin{align}
P& =b_{1}^{2}\cos ^{2}\frac{\alpha }{2}+b_{2}^{2}\sin ^{2}\frac{\alpha }{2}%
+a_{1}^{2},  \notag \\
Q& =b_{1}^{2}\sin ^{2}\frac{\alpha }{2}+b_{2}^{2}\cos ^{2}\frac{\alpha }{2}%
+a_{2}^{2}, \\
R& =2\left[ b_{2}^{2}\cos \frac{\alpha }{2}\sin \frac{\alpha }{2}%
-b_{1}^{2}\cos \frac{\alpha }{2}\sin \frac{\alpha }{2}\right] .  \notag
\end{align}%
Therefore, the integral
\begin{align*}
& \int \int \exp \left\{ -\frac{1}{2}\left(
a_{1}^{2}x_{1}^{2}+a_{2}^{2}x_{2}^{2}+b_{1}^{2}y_{1}^{2}+b_{2}^{2}y_{2}^{2}%
\right) \right\} \\
& \times H_{n}\left( a_{1}x_{1}\right) H_{0}\left( a_{2}x_{2}\right)
H_{n}\left( b_{1}y_{1}\right) H_{0}\left( b_{2}y_{2}\right) dx_{1}dx_{2}\,
\end{align*}%
is transformed to
\begin{align}
& \int \int \exp \left\{ -\frac{1}{2}W_{1}z_{1}^{2}-\frac{1}{2}%
W_{2}z_{2}^{2}\right\}  \notag \\
& \times H_{n}\left( a_{1}x_{1}\right) H_{0}\left( a_{2}x_{2}\right)
H_{n}\left( b_{1}y_{1}\right) H_{0}\left( b_{2}y_{2}\right) Jdz_{1}dz_{2}
\notag \\
&  \label{integral_z}
\end{align}%
where%
\begin{equation*}
J=\left\vert
\begin{array}{cc}
\frac{\partial x_{1}}{\partial z_{1}} & \frac{\partial x_{1}}{\partial z_{2}}
\\
\frac{\partial x_{2}}{\partial z_{1}} & \frac{\partial x_{2}}{\partial z_{2}}%
\end{array}%
\right\vert =1
\end{equation*}%
is the Jacobi determinant and
\begin{align}
W_{1}& =\frac{P+Q-S}{2},  \notag \\
W_{2}& =\frac{P+Q+S}{2},
\end{align}%
where
\begin{equation*}
S=\left( Q-P\right) \sqrt{1+\frac{R^{2}}{\left( Q-P\right) ^{2}}}
\end{equation*}%
and

\begin{equation*}
\left(
\begin{array}{c}
x_{1} \\
x_{2}%
\end{array}%
\right) =\left(
\begin{array}{cc}
\cos \frac{\beta }{2} & \sin \frac{\beta }{2} \\
-\sin \frac{\beta }{2} & \cos \frac{\beta }{2}%
\end{array}%
\right) \left(
\begin{array}{c}
z_{1} \\
z_{2}%
\end{array}%
\right) ,
\end{equation*}%
\begin{equation*}
\left(
\begin{array}{c}
y_{1} \\
y_{2}%
\end{array}%
\right) =\left(
\begin{array}{cc}
\cos \frac{\alpha -\beta }{2} & -\sin \frac{\alpha -\beta }{2} \\
\sin \frac{\alpha -\beta }{2} & \cos \frac{\alpha -\beta }{2}%
\end{array}%
\right) \left(
\begin{array}{c}
z_{1} \\
z_{2}%
\end{array}%
\right) .
\end{equation*}%
We note that $H_{n}\left( b_{1}y_{1}\right) ,\,H_{0}\left( b_{2}y_{2}\right)
,\,H_{n}\left( a_{1}x_{1}\right) ,\,H_{0}\left( a_{2}x_{2}\right) $ can be
expressed as a polynomial of $z_{1}\,$and$\,z_{2}$, so that the integral (%
\ref{integral_z}) can be calculated with the help of the following integral
formula
\begin{equation*}
\int_{-\infty }^{+\infty }t^{n}e^{-\rho t^{2}}=\frac{1}{2}\left( 1+\left(
-1\right) ^{n}\right) \rho ^{-\frac{1}{2}\left( 1+n\right) }\Gamma \left[
\frac{1+n}{2}\right] ,
\end{equation*}%
where $n=0,\,1,\,\cdots \,,\,\rho >0,$ and $\Gamma \lbrack n]$ is the gamma
function.

For the simplest case, when $n=1$, we have%
\begin{align}
F_{1}=\left\vert \frac{4}{\pi }\sqrt{\frac{a_{1}^{3}a_{2}b_{1}^{3}b_{2}}{%
W_{1}W_{2}}}\times \right. & \left[ \frac{1}{W_{1}}\cos \frac{\alpha -\beta
}{2}\cos \frac{\beta }{2}\right.  \notag \\
& \left. \left. -\frac{1}{W_{2}}\sin \frac{\alpha -\beta }{2}\sin \frac{%
\beta }{2}\right] \right\vert ^{2}\,.
\end{align}

\end{document}